\documentclass[12pt,a4paper]{article}
\pdfoutput=1

\usepackage{jheppub}
\usepackage{amsmath,amsfonts,mathrsfs,graphicx,bbm}
\usepackage{fixltx2e}

\usepackage[font=small]{caption}

\usepackage{lmodern}

\usepackage{graphicx}
\usepackage{booktabs}
\usepackage{dsfont}

\usepackage{mathrsfs}
\usepackage{amsmath,amssymb}
\usepackage{mathtools}
\usepackage{bbm}
\usepackage{slashed}
\usepackage{braket}

\usepackage{hyperref}

\usepackage{xspace}

\usepackage{color}

\definecolor{grey}{rgb}{0.4,0.4,0.5}
\definecolor{darkgreen}{rgb}{0,0.5,0}
\definecolor{darkred}{rgb}{0.6,0.0,0}
\definecolor{lightbrown}{rgb}{1,0.9,0.8}
\definecolor{brown}{rgb}{0.6,0.3,0.3}
\definecolor{darkblue}{rgb}{0,0,0.8}
\definecolor{darkmagenta}{rgb}{0.5,0,0.5}

\def\de{\delta }

\numberwithin{equation}{section}

\makeatletter
 \let\old@startsection=\@startsection
 \let\oldl@section=\l@section
 \renewcommand{\@startsection}[6]{\old@startsection{#1}{#2}{#3}{#4}{#5}{#6\mathversion{bold}}}
 \renewcommand{\l@section}[2]{\oldl@section{\mathversion{bold}#1}{#2}}
\makeatother

\def\XXint#1#2#3{{\setbox0=\hbox{$#1{#2#3}{\int}$}
    \vcenter{\hbox{$#2#3$}}\kern-.5\wd0}}

\newcommand{\alg}[1]{\mathfrak{#1}}

\newcommand{\be}{\begin{equation}}
\newcommand{\ee}{\end{equation}}
\newcommand{\bea}{\begin{eqnarray}}
\newcommand{\eea}{\end{eqnarray}}
\newcommand{\bal}{\begin{equation}\begin{aligned}}
\newcommand{\eal}{\end{aligned}\end{equation}}

\newcommand{\bee}{\begin{enumerate}}
\newcommand{\eee}{\end{enumerate}}
\newcommand{\bei}{\begin{itemize}}
\newcommand{\eei}{\end{itemize}}

\def\ov{\over}

\def\la{\label}

\def\a {\alpha}
\def\b {\beta}

\def\g {\gamma}

\newcommand{\sfrac}[2]{{\textstyle\frac{#1}{#2}}}

\def\x'{\mathaccent 19 x}
\def\y'{\mathaccent 19 y}
\def\n'{\mathaccent 19 n}
\def\u'{\mathaccent 19 u}

\def\et'{\mathaccent 19 \eta}
\def\th'{\mathaccent 19 \theta}
\def\lam'{\mathaccent 19 \lambda}
\def\varet'{\mathaccent 19 \vartheta}
\def\rh'{\mathaccent 19 \rho}
\def\ph'{\mathaccent 19 \phi}
\def\xb'{\mathaccent 19 {\bar{x}}}

\def\na{{\nabla}}

\def\sl(2){\alg{sl}(2)}

\def\be{\begin{equation}}
\def\ee{\end{equation}}

\def\a {\alpha}
\def\b {\beta}

\def\g {\gamma}

\def\la{\label}

\def\ov{\over}

\def\d1{{\dot{1}}}

\newcommand{\bem}{\left (\begin{matrix}}
\newcommand{\eem}{\end{matrix} \right )}

\title{Towards 4-point correlation functions of \\
any $\sfrac{1}{2}$-BPS operators from supergravity}

\author[a,1]{Gleb Arutyunov,}
\note{Correspondent fellow at Steklov
Mathematical Institute, Moscow.}
\author[b,1]{Sergey Frolov,}
\author[a]{Rob Klabbers}
\author[a]{and Sergei Savin}
\affiliation[a]{II. Institut f\"ur Theoretische Physik, Universit\"at Hamburg, Luruper Chaussee 149, 22761 Hamburg, Germany\\
Zentrum f\"ur Mathematische Physik, Universit\"at Hamburg, Bundesstrasse 55, 20146 Hamburg, Germany
}
\affiliation[b]{Hamilton Mathematics Institute and School of Mathematics, \\
~~Trinity College, Dublin 2, Ireland}
\emailAdd{gleb.arutyunov@desy.de}  
\emailAdd{frolovs@maths.tcd.ie}
\emailAdd {rob.klabbers@desy.de}
\emailAdd{spsavin@gmail.com}

\abstract{ The quartic effective action for Kaluza-Klein modes that arises upon compactification of type IIB supergravity on the five-sphere ${\rm S}^5$ is a starting point for computing 
the four-point correlation functions of arbitrary weight $\sfrac{1}{2}$-BPS operators in ${\cal N}=4$ super Yang-Mills theory in the supergravity approximation.
The apparent structure of this action is rather involved, in particular  it contains quartic terms with four derivatives which cannot be removed by field redefinitions.  
By exhibiting intricate identities between certain integrals involving spherical harmonics of ${\rm S}^5$ we show that the net contribution of these four-derivative terms to the 
effective action vanishes.  Our result is in agreement with and provides further support to the recent conjecture on the Mellin space representation of the four-point correlation function of any 
$\sfrac{1}{2}$-BPS operators in the supergravity approximation. 

}

\begin{document}

\begin{flushright}
\scriptsize{TCD-MATH-17-01\\
ZMP-HH-17-01}
\end{flushright}

\maketitle
\flushbottom

\renewcommand{\thefootnote}{\arabic{footnote}}
\setcounter{footnote}{0}

\section{Introduction and summary}
Recently the problem of finding four-point correlation functions of $\sfrac{1}{2}$-BPS operators in ${\cal N}=4$ super Yang-Mills theory 
came into focus again. Both new demands and inspirations sparked due to interesting progress
of the general bootstrap program, as well as remarkable developments concerning the integrable structure of the dual string theory. The main 
interest in these correlation functions  is that on the one hand they involve the simplest possible operators, 
while on the other hand they non-trivially depend on the coupling constant and, as such, contain through the corresponding OPEs a lot of valuable information 
about the dynamics of unprotected operators. 

It was known for some time  \cite{Mack:2009mi}-\cite{Fitzpatrick:2011ia} that the simplest representation of the correlation functions is achieved by transforming the latter to  Mellin space,
where their analytic and asymptotic properties become most transparent. 
Recently the authors of  \cite{Rastelli:2016nze}   conjectured an interesting formula for 
the Mellin representation of the four-point correlation function of arbitrary weight $\sfrac{1}{2}$-BPS operators  in the supergravity limit. 
In addition to the expected OPE behavior in this limit, the conjecture is based on the assumption of  linear asymptotic growth of the Mellin amplitude  
at large values of the Mandelstam variables. 

Leaving technical complications out of the  discussion, to derive the corresponding correlation function  and in this way to prove the conjecture by  \cite{Rastelli:2016nze},  one has to use 
the quartic effective action for Kaluza-Klein modes of type IIB supergravity compactified on the five-sphere ${\rm S}^5$ \cite{Arutyunov:1999fb}.  
We recall that $\sfrac{1}{2}$-BPS operators of weight $k$ are dual to supergravity scalars $s^{I_k}$ with mass $m^2=k(k-4)$ and with index $I_k$ 
running over the basis of an irreducible representation $[0,k,0]$ 
of ${\rm SU(4)}$. Then a four-point correlation function ${\cal A}$  comes naturally as a sum of two terms  
\bea
\nonumber
{\cal A}={\cal A}_{\rm exchange}+{\cal A}_{\rm contact}\, .
\eea
Here ${\cal A}_{\rm exchange}$ comprises the sum of all exchange graphs which are generated by cubic Lagrangian vertices involving two scalars $s^I$ and any other supergravity 
field which is allowed by representation theory. The term ${\cal A}_{\rm contact}$  gives a contribution of all contact graphs originating from the quartic Lagrangian vertices, {\it i.e.}
the vertices which contain four scalars $s^I$ possibly with space-time derivatives.  

Analysing the behaviour of exchange graphs for large values of Mandelstam variables one concludes that they grow linearly, hence the conjecture by  \cite{Rastelli:2016nze}.
As to the contact graphs, their linear asymptotic growth is guaranteed provided the quartic Lagrangian vertices contain fields $s^I$ with a number of derivatives not higher than two.
Puzzling enough, such an  expected structure of the quartic Lagrangian appears in apparent conflict with the explicit findings by  \cite{Arutyunov:1999fb}, where it was 
shown that the quartic Lagrangian contains terms with four derivatives (and not more) and that these four-derivative terms cannot be removed by any on-shell field redefinition.
When genuinely present, such terms would result into quadratic asymptotic growth of the Mellin amplitude, incompatible with the conjectured formula.    

Obviously, the arising conundrum has a natural resolution if the four-derivative couplings actually vanish due to some internal, group-theoretic reasons. 
The idea that this might be the case, which implies that the corresponding Kaluza-Klein effective action is of the sigma-model type, 
has been put forward long ago\footnote{See \cite{Arutyunov:2002fh}, the discussion around formula (4.4) there.}. It is essentially based on the results of evaluating
concrete correlation functions. Namely,  using the quartic effective action, correlation functions of weight $k$ BPS operators of the type $\langle kkkk\rangle$
and $\langle 22kk\rangle$ were computed  \cite{Arutyunov:2002fh}-\cite{Uruchurtu:2011wh}, and in all the cases one was able to demonstrate vanishing of  the quartic 
four-derivative terms by integrating by parts together with specifying  explicit group-theoretic content of the corresponding Lagrangian couplings. 
Unfortunately, these correlators are still rather special and they do not probe all quartic four-derivative couplings to draw a decisive conclusion on the ultimate status of the latter.  

In this note we show that the quartic four-derivative terms do indeed vanish. On the one hand, this gives evidence in favour of the conjecture  \cite{Rastelli:2016nze}
and, on the other hand, provides a first step towards actual evaluation of the four-point correlation function of arbitrary weight $\sfrac{1}{2}$-BPS operators in the supergravity approximation.

Our considerations are based on the fact that in the process of compactification the couplings in front of quartic terms with four derivatives 
appear in the form of weighted sums of two Clebsch-Gordan coefficients $c$ which look schematically as $g_5c_{125}c_{345}$. Here the legs $1,2,3,4$ carry representation indices  of 
supergravity field $s^I$ with weights $k_1$, $k_2$, $k_3$ and $k_4$, while summation over the fifth leg with some weight function $g_5$
is assumed.  The Clebsch-Gordan coefficients $c_{125}$ are given in terms of integrals of spherical harmonics of ${\rm S}^5$, where
in the fifth place occurs either a scalar harmonic $Y^{I_5}$ or a vector one   $Y_{\a}^{I_5}$, where $\a$ is the tangent index of the five-sphere. By unravelling 
some intricate identity which involves a sum of integrals over vector harmonics, we managed to rewrite the contribution of the latter 
via similar sums but involving scalar harmonics only.   By removing vector harmonics in favour of scalar ones in this way, we made all the couplings comparable,
and further summing them up we find that they exactly cancel.

Although our result on vanishing of quartic four-derivative terms provides a rigorous proof that the quartic effective action for Kaluza-Klein modes of type IIB supergravity is of the sigma-model 
type, a deep reason behind this finding still remains to be understood.  We point out that similar manipulations might exist for the rest of the quartic effective action, which comprises 
terms with two derivatives and without derivatives, possibly leading to its significant simplification.  
We however postpone investigation of this interesting question for future work.

The paper is organised as follows. In the next section we present the known information about quartic couplings with four derivatives and collect all necessary 
definitions. In Section 3 we derive the main reduction formula which allows us to replace the coupling with vector spherical harmonics via new couplings with scalar harmonics only,
and verify that upon this replacement the net contribution of all four-derivative couplings sums up to zero. A  computation of an auxiliary integral is relegated to an appendix.

\section{Quartic couplings with four derivatives}
In \cite{Arutyunov:1999fb} the quartic Lagrangian for the fields $s^I$ with four derivatives was found to be of the following form
\bea
\label{Lag}
{\cal L}_4^{(4)}=\sum_{1,2,3,4}\big(S_{1234}^{(4)}+A_{1234}^{(4)}\big)s^{1}\nabla_a s^{2}\nabla_b^2(s^{3}\nabla^a s^{4})\, .
\eea
Here $\nabla_a$  is a covariant derivative along AdS space and each summation label $j=1,\ldots, 4$ stands for a concise notation for the representation index $I_j$ running over a basis of of an irreducible representation $[0,k_j,0]$ of ${\rm SU(4)}$. 
The couplings $A_{1234}$ and $S_{1234}$ have the following symmetry properties
\begin{equation}
\label{AS}
\begin{aligned}
A_{1234}^{(4)}&=-A_{2134}^{(4)}=A_{3412}^{(4)}\, , \\
S_{1234}^{(4)}&=S_{2134}^{(4)}=S_{3412}^{(4)}\, .
\end{aligned}
\end{equation}
Explicitly, $A_{1234}$ is given by the sum of the following individual terms
\bea
(A_3)^{(4)}_{1234}&=&\frac{1}{4\de}
f_5^3\left( a_{145}a_{235}-a_{135}a_{245}\right) .\nonumber\\
(A_2)^{(4)}_{1234}&=&-\frac{1}{4\de}(3(f_1+f_2+f_3+f_4)-28)
f_5^2\left( a_{145}a_{235}-a_{135}a_{245}\right) .\nonumber\\
(A_1)^{(4)}_{1234}&=&-\frac{3}{4\de}(f_1-f_2)(f_3-f_4)
f_5 a_{125}a_{345} \nonumber\\
&+&\frac{1}{2\de}(f_1+f_2+f_3+f_4-2)(f_1+f_2+f_3+f_4-12)
f_5\left( a_{145}a_{235}-a_{135}a_{245}\right) .\nonumber
\eea

\bea
(A_0)^{(4)}_{1234}&=&\frac{21}{4\de}(f_1-f_2)(f_3-f_4)
 a_{125}a_{345}.\nonumber\\
(A_{-1})^{(4)}_{1234}&=&-\frac{12}{\de}(f_1-f_2)(f_3-f_4)
 f_5^{-1}a_{125}a_{345}.\nonumber\\
(A_{t2})^{(4)}_{1234}&=&-\frac{3}{\de}(f_5-1)^2t_{125}t_{345}.
\nonumber
\eea
The symmetric coupling is 
\bea
\nonumber
S_{1234}^{(4)}&=&
\frac{7}{4\de}\left( 2f_1f_2+2f_3f_4-(f_1+f_2)(f_3+f_4)\right)
 a_{125}a_{345}.\nonumber
\eea
In the formulae above $f_i\equiv f(k_i)=k_i(k_i+4)$, $\delta=\prod_{i=1}^4(k_i+1)$ and summation over the index $5$ 
is assumed. The couplings $a_{123}$ and $t_{123}$ are given as the following integrals over the five-sphere of the spherical harmonics 
\bea
a_{123}=\int Y^1Y^2Y^3\, , ~~~~~~~~~t_{123}=\int \nabla^{\a} Y^1 Y^2 Y^3_{\alpha}\, .
\eea
Here $Y^k$ are scalar spherical harmonics and $Y^k_{\a}$ are vector spherical harmonics satisfying the irreducibility condition $\nabla^{\a}Y_{\a}^k=0$. 
Both $Y^k$ and $Y_{\a}^k$ are eigenvalues of the sphere Laplacian $\nabla^2$ with the following eigenvalues
\bea
\la{Laplacian}
\nabla^2Y^k=-f_k Y^k\, ,~~~~~~~\nabla^2Y_{\alpha}^k=(1-f_k)Y_{\alpha}^k\, . 
\eea 
In what follows we will also need the following product formulae which follow from the orthogonality relation for scalar harmonics
\bea
Y^1Y^2=a_{125}Y^5,\quad \na^\a Y^1\na_\a Y^2=b_{125}Y^5,\quad
\na^\a\na^\b Y^1\na_\a\na_\b Y^2=c_{125}Y^5\,,
\la{rel1}
\eea
where the coefficients are\footnote{The formula for $c_{123}$ in terms of $a_{123}$ is different from the one in \cite{Arutyunov:1999fb}, because there
the combination $ \na^{\a}\na^{\b}$ stands for the traceless symmetric combination of derivatives $ \na^{\a}\na^{\b}\equiv \na^{(\a}\na^{\b)}$.}
\bal
\la{bc}
b_{123}&=\int \na^{\a}Y^{1}\na_{\a}Y^{2}Y^{3}
=\frac{1}{2}(f_1+f_2-f_3)a_{123}, \\
c_{123}&=\int \na^{\a}\na^{\b}Y^{1}\na_{\a}\na_{\b}Y^{2}Y^{3}=
\frac{1}{2}(f_1+f_2-f_3-8)(f_1+f_2-f_3) a_{123}\, .
\eal
This completes our discussion of the known results on the quartic Lagrangian with four-derivative vertices, for further information and derivation of the above formulae 
we refer the reader to \cite{Arutyunov:1999fb}.

To proceed with proving our main result, we employ the same strategy as in \cite{Arutyunov:2000ima}, where the vanishing of quartic four-derivative-vertices were shown for the so-called sub-extremal
and sub-sub-extremal cases. Recall that we are ultimately interested in the four-point function of BPS operators corresponding to arbitrary weights $k_1,\ldots k_4$. We can therefore 
restrict the infinite sum in (\ref{Lag}) to representations which correspond to these weights. The sum in (\ref{Lag}) is not ordered and, therefore, there are 24 ordered sets of 
the indices $k_1,\ldots, k_4$ which split into 3 equivalence classes due to the symmetries  (\ref{AS}). Further, integrating by parts and using (\ref{AS}), we represent the part of the Lagrangian (\ref{Lag}) 
contributing to the four-point function $\langle k_1k_2k_3k_4\rangle$ in the form similar to that in \cite{Arutyunov:2000ima}
\bea
{\cal L}_4^{(4),k_1k_2k_3k_4}=&-&8\sum_{1,2,3,4}\Big( S_{1234}^{(4)}
+A_{1324}^{(4)}+A_{1423}^{(4)}\Big)
\na_a s^{1}\na^as^{2}\na_bs^{3}\na^bs^{4}\, \nonumber \\
&-&8\sum_{1,2,3,4}\Big( S_{1324}^{(4)}
+A_{1234}^{(4)}+A_{1432}^{(4)}\Big)
\na_a s^{1}\na^as^{3}\na_bs^{2}\na^bs^{4}    \la{4}  \\
&-&8\sum_{1,2,3,4}\Big( S_{1432}^{(4)}
+A_{1342}^{(4)}+A_{1243}^{(4)}\Big)
\na_a s^{1}\na^as^{4}\na_bs^{2}\na^bs^{3}\, . \nonumber \eea
Since we are interested here in the four-derivative vertices only, in the above formula 
we have omitted the contribution of two-derivative terms and terms without derivatives which arise upon integrating by parts and 
using equations of motion. There terms however should be taken into account in subsequent analysis of the remaining part of the quartic effective action.
We also note that the meaning of the sums in (\ref{4}) is different from that in (\ref{Lag}) -- in (\ref{4}) the sums are ordered, {\it i.e.} summation over $1$ means summation over 
index $I_1$ corresponding to the representation with a given weight $k_1$ and so on. It is now obvious that it is enough to analyse the coupling
\bea
\la{main_coupling}
{\mathscr C}_{1234}\equiv S_{1234}^{(4)} + A_{1324}^{(4)} + A_{1423}^{(4)} \, ,
\eea
because the other two couplings in (\ref{4}) differ from it by permutation of indices only.

Obviously, among the couplings there is a distinguished one, namely, $(A_{t2})^{(4)}_{1234}$, as the latter involves vector spherical harmonics. Its contribution into (\ref{main_coupling})
comes in the combination
\bea
\la{need_red}
W^{1234}\equiv (f_5-1)^2(t_{135}t_{245}+t_{145}t_{235})\, .
\eea
Our further strategy will be to reduce this combination to structures of the type $f_5^na_{125}a_{345}$ and permutations thereof. After this is done,
all the couplings become comparable and we can add them up according to (\ref{main_coupling}). 

\section{Reduction formula}
The reduction of (\ref{need_red}) is based on the following formula  \cite{Arutyunov:1999fb}
\bal
\na_{\a} Y^1 Y^2=t_{125}Y_{\a}^5+\frac{b_{152}}{f_5}\na_{\a }Y^5 \, .
\la{vecdec}
\eal
In what follows it appears advantageous to split (\ref{vecdec}) into anti-symmetric and symmetric part with respect to indices $1$ and $2$, namely,
\bea
\na_{\a} Y^{[1} Y^{2]}&\equiv &{1\ov2}(\na_{\a} Y^1 Y^2-\na_{\a} Y^2 Y^1)= t_{125}Y_{\a}^5+\frac{f_{1}-f_{2}}{2f_5}a_{125}\na_{\a }Y^5\,, 
\la{vecdec2} \\
\na_{\a} (Y^1 Y^2)&=&a_{125}\na_{\a }Y^5 \, .
\la{vecdec2s}
\eea
Acting on (\ref{vecdec2}) with the Laplacian and taking into account that 
\bal
\la{LspY}
\na^2 \na_\a Y^k=-(f_k-4)\na_\a Y^k\, ,
\eal
we obtain
\bal
(1-f_5)t_{125}Y_{\a}^5=\na^2(\na_{\a} Y^{[1} Y^{2]})-\frac{f_{1}-f_{2}}{2f_5}(4-f_5)a_{125}\na_{\a }Y^5\, .
\la{vecdec34}
\eal
Now we multiply this relation with a similar one where indices $1,2,5$ are replaced by $3,4,6$ and integrate over the five-sphere. The orthogonality relation for the vector spherical harmonics 
together with (\ref{LspY}) used upon integrating the Laplacian by parts leads to the following formula 
\bal
(f_5-1)^2t_{125}t_{345}&=\int \na^2(\na_{\a}Y^{[1}Y^{2]})\na^2(\na^\a Y^{[3} Y^{4]})- \nonumber\\
&~~~~~~~~~~~~~~~-\frac{(4-f_5)^2}{4f_5}(f_1-f_2)(f_3-f_4)a_{125}a_{345} \, ,
\eal
which for (\ref{need_red}) implies the following relation
\bal
W^{1234}=K^{1234}-\frac{(4-f_5)^2}{4f_5}\Big((f_1-f_3)(f_2-f_4)a_{135}a_{245} +(f_1-f_4)(f_2-f_3)a_{145}a_{235}\Big)\, , \nonumber
\eal
where $K^{1234}$ is the following integral
\bal
\la{K}
K^{1234}&=\int \left( \na^2(\na_{\a} Y^{[1} Y^{4]})\na^2(\na_{\a} Y^{[2} Y^{3]}) + \na^2(\na_{\a} Y^{[1} Y^{3]})\na^2(\na_{\a} Y^{[2} Y^{4]}) \right)
\eal
with the symmetry properties
\bal
K^{1234}=K^{2134}=K^{1243}=K^{3412}\, .
\eal
Thus, we have reduced evaluation of our main quantity $W^{1234}$ to the computation of the integral $K^{1234}$. In order not to overload our discussion with heavy formulae 
we perform the computation of $K^{1234}$ in the appendix and here present only the final reduction formula for $W^{1234}$:
{\small
\bea
 &&W^{1234}= \nonumber\\
 && -\frac{a_{135} a_{245} }{24
   f_5}\Big(-2 f_5^4+2 (3 f_1+3 f_2+3 f_3+3 f_4-28) f_5^3-2 (2
   f_1^2-(f_2-5 f_3-8 f_4+28) f_1 \nonumber \\
   &&+2 f_2^2+2 f_4^2+2 (f_3-12)
   (f_3-2)-(f_3+28) f_4+f_2 (8 f_3+5 f_4-28))
   f_5^2  \nonumber  \\
   &&+((f_2+3 f_3+4 f_4-12) f_1^2+(f_2^2+4 (f_3+f_4-20) f_2+3
   (f_3-4)^2+4 f_4^2+4 (f_3+4) f_4) f_1  \nonumber  \\
   &&+(3 f_2+f_3-12)
   f_4^2+4 ((f_3-3) f_2^2+(f_3 (f_3+4)+12) f_2-3
   (f_3-4) f_3)+(3 (f_2-4)^2 \nonumber  \\
   &&+f_3^2+4 (f_2-20)
   f_3) f_4) f_5+96 (f_1-f_3) (f_2-f_4)\Big)   \nonumber \\
   &&-\frac{a_{145} a_{235}}{24 f_5} \Big(-2 f_5^4+2 (3 f_1+3 f_2+3 f_3+3 f_4-28) f_5^3-2
   (2 f_1^2-(f_2-8 f_3-5 f_4+28) f_1 \nonumber \\
   &&+2 f_2^2+2 f_4^2+2 (f_3-12)
   (f_3-2)-(f_3+28) f_4+f_2 (5 f_3+8 f_4-28))
   f_5^2 \nonumber \\
   &&+((f_2+4 f_3+3 (f_4-4)) f_1^2+(f_2^2+4
   (f_3+f_4-20) f_2+4 f_3^2+3 (f_4-4)^2+4 f_3 (f_4+4))
   f_1 \nonumber \\
   &&+(4 f_2+f_3-12) f_4^2+3 (f_2-4) (f_3-4)
   (f_2+f_3) \nonumber \\
   &&+(4 f_2^2+4 (f_3+4) f_2+(f_3-80) f_3+48)
   f_4) f_5+96 (f_2-f_3) (f_1-f_4)\Big) \nonumber \\
  &&+\frac{  a_{125} a_{345}}{24}
  \Big(-4 f_5^3+4 (3 f_3+3 f_4-28) f_5^2-4 (2 f_4^2+5 f_3
   f_4-28 f_4+2 (f_3-14) f_3+48) f_5 \nonumber \\
   &&+6 (f_3-4) (f_4-4)
   (f_3+f_4)+f_2 (5 f_3^2+8 f_4 f_3-64 f_3+5 f_4^2+12 f_5^2-64 f_4 \nonumber \\
   &&-14
   (f_3+f_4-8) f_5+96)+f_1 (6 f_2^2+4 (2 (f_3+f_4-6)-5
   f_5) f_2+5 f_3^2+5 f_4^2+12 f_5^2 \nonumber \\
   &&-64 f_3+8 f_3 f_4-64 f_4-14 (f_3+f_4-8)
   f_5+96)+f_2^2 (5 f_3+5 f_4-8 (f_5+3)) \nonumber \\
   &&+f_1^2 (6 f_2+5 f_3+5
   f_4-8 (f_5+3))\Big)\, .
    \la{W}
 \eea
}
\normalsize

\vskip -0.6cm
\noindent
Finally, we sum $-\frac{3}{\delta}W^{1234}$ with the remaining couplings in (\ref{main_coupling}) and observe
that all the terms are neatly cancelled delivering thereby ${\mathscr C}_{1234}=0$. In this way we have shown that 
the quartic four-derivative couplings of the supergravity effective action vanish.

\section{Acknowledgements}
G.A. and S.F. would like to thank Leonardo Rastelli for the revival of our interest in the supergravity action for Kaluza-Klein modes  and for useful discussions. 
The work of G.A., R.K. and S.S. is supported by the German Science Foundation (DFG) 
under the Collaborative Research Center (SFB) 676 Particles, Strings and the Early Universe and the Research Training Group 1670.

\appendix

\section{Evaluation of $K^{1234}$}
In this appendix we compute the integral (\ref{K}) in terms of structures $f_5^na_{125}a_{345}$ and permutations thereof. 
In the computation process the following identity
valid for any co-vector $\xi_\a$ 
\bal
\la{comm}
 \left[\na_\a\,,\, \na_\b\right] \xi_\g=g_{\a\g}\xi_\b-g_{\b\g}\xi_\a\,.
\eal
will be heavily used. Here $g_{\a\beta}$ is the metric of the unit five-sphere. Note also that on a scalar function two covariant derivatives commute.
\smallskip

\noindent
We start with computing
\bea
 \na^2(\na_{\a} Y^1 Y^4)
 &=&2\na_{\b}\na_{\a} Y^1 \na^{\b} Y^4-(f_1+f_4-4)\na_{\a} Y^1 Y^4 \, .
\eea
The latter formula gives rise to the following identity
\bea\la{Id}
 &&\na^2(\na_{\a} Y^1 Y^4)\na^2(\na^{\a} Y^2 Y^3)=4\na_{\b}\na_{\a} Y^1 \na^{\b} Y^4\na_{\g}\na^{\a} Y^2 \na^{\g} Y^3  \\
 &&~~~-2(f_2+f_3-4)\na_{\b}\na_{\a} Y^1 \na^{\b} Y^4\na^{\a} Y^2 Y^3 - 2(f_1+f_4-4)\na_{\a} Y^1 Y^4\na_{\g}\na^{\a} Y^2 \na^{\g} Y^3 \nonumber\\
 &&~~~+(f_1+f_4-4)(f_2+f_3-4)\na_{\a} Y^1 Y^4\na^{\a} Y^2 Y^3  \nonumber
 \, 
\eea
and a similar one with indices $3$ and $4$ interchanged. To simplify our presentation, 
in the sequel we will drop the integration sign and always identify expressions differing by a total derivative. Using (\ref{Id})  we then get
\bea\la{K1234}
K^{1234}&=& U^{1234}+V^{1234}\,,
 \eea
 where 
 \bea
&&U^{1234}=\\
&&-2(f_2+f_3-4)\na_{\b}\na_{\a} Y^{[1} \na^{\b} Y^{4]}\na^{\a} Y^{[2} Y^{3]}-2(f_2+f_4-4)\na_{\b}\na_{\a} Y^{[1} \na^{\b} Y^{3]}\na^{\a} Y^{[2} Y^{4]} \nonumber\\
&&-2(f_1+f_4-4)\na_{\a} Y^{[1} Y^{4]}\na_{\g}\na^{\a} Y^{[2} \na^{\g} Y^{3]}-2(f_1+f_3-4)\na_{\a} Y^{[1} Y^{3]}\na_{\g}\na^{\a} Y^{[2} \na^{\g} Y^{4]} \nonumber\\
&&+ 
 (f_1+f_4-4)(f_2+f_3-4)\na_{\a} Y^{[1} Y^{4]}\na^{\a} Y^{[2} Y^{3]} \nonumber \\
&&+ 
 (f_1+f_3-4)(f_2+f_4-4)\na_{\a} Y^{[1} Y^{3]}\na^{\a} Y^{[2} Y^{4]}\, .  \nonumber
 \eea
 and
\bea
\hspace{-0.8cm}
V^{1234}=4\big(\na_{\b}\na_{\a} Y^{[1} \na^{\b} Y^{4]}\na_{\g}\na^{\a} Y^{[2} \na^{\g} Y^{3]} +\na_{\b}\na_{\a} Y^{[1} \na^{\b} Y^{3]}\na_{\g}\na^{\a} Y^{[2} \na^{\g} Y^{4]}\big)\, .\la{V1234}
\eea
We continue our further treatment with evaluating the quantity $U^{1234}$. To this end, we compute
 \bea
&&\na_{\b}\na_{\a} Y^{[i} \na^{\b} Y^{j]}\na^{\a} Y^{[k} Y^{l]} \nonumber \\
&&=-\na_{\a} Y^{[i} \na^2 Y^{j]}\na^{\a} Y^{[k} Y^{l]}-\na_{\a} Y^{[i} \na_{\b} Y^{j]}\na^{\b}\na^{\a} Y^{[k} Y^{l]}-\na_{\a} Y^{[i} \na_{\b} Y^{j]}\na^{\a} Y^{[k} \na^{\b}Y^{l]} \nonumber \\
&&={1\ov4}(f_j\na_{\a} Y^{i} Y^{j}-f_i\na_{\a} Y^{j} Y^{i})(\na^{\a} Y^{k} Y^{l}-\na^{\a} Y^{l} Y^{k}) \nonumber \\
&&-{1\ov4}(\na_{\a} Y^{i} \na_{\b} Y^{j}-\na_{\a} Y^{j} \na_{\b} Y^{i})(\na^{\a} Y^{k} \na^{\b}Y^{l}-\na^{\a} Y^{l} \na^{\b}Y^{k}) \la{Yijkl1}  \\
&&={1\ov4}f_j(b_{ik5}a_{jl5} -b_{il5}a_{jk5})-{1\ov4}f_i(b_{jk5}a_{il5} -b_{jl5}a_{ik5})-{1\ov2}(b_{ik5}b_{jl5} -b_{il5}b_{jk5}) \nonumber \\
&&=\frac{1}{8} ((f_5-f_k) (f_5-f_l)-f_i f_j) (a_{il5} a_{jk5}-a_{ik5} a_{jl5}) \nonumber
\,,
 \eea
and
\bal\la{Yijkl2}
&\na_{\a} Y^{[i} Y^{j]}\na^{\a} Y^{[k} Y^{l]}={1\ov4}(\na_{\a} Y^{i} Y^{j}-\na_{\a} Y^{j} Y^{i})(\na^{\a} Y^{k} Y^{l}-\na^{\a} Y^{l} Y^{k})\\
&={1\ov4}(b_{ik5}a_{jl5} -b_{il5}a_{jk5})-{1\ov4}(b_{jk5}a_{il5} -b_{jl5}a_{ik5})\\
&=\frac{1}{8} (-f_i-f_j-f_k-f_l+2 f_5) (a_{il5} a_{jk5}-a_{ik5} a_{jl5})
\, .
 \eal
As the result, for the quantity $U^{1234}$ we get
 \bal
 U^{1234}&=\frac{1}{8} (-2 (f_{1}+f_{4}-4) ((f_{5}-f_{1})
   (f_{5}-f_{4})-f_{2} f_{3})\\
   &-(f_{1}+f_{4}-4)
   (f_{2}+f_{3}-4) (f_{1}+f_{2}+f_{3}+f_{4}-2
   f_{5})\\
   &-2 (f_{2}+f_{3}-4) ((f_{5}-f_{2})
   (f_{5}-f_{3})-f_{1} f_{4})) (a_{{1}{3}5}
   a_{{2}{4}5}-a_{{1}{2}5}
   a_{{3}{4}5})\\
   &+\frac{1}{8} (-2 (f_{2}+f_{4}-4)
   ((f_{5}-f_{2}) (f_{5}-f_{4})-f_{1}
   f_{3})\\
   &-(f_{1}+f_{3}-4) (f_{2}+f_{4}-4)
   (f_{1}+f_{2}+f_{3}+f_{4}-2 f_{5})\\
   &-2
   (f_{1}+f_{3}-4) ((f_{5}-f_{1}) (f_{5}-f_{3})-f_{2}
   f_{4})) (a_{{1}{4}5}
   a_{{2}{3}5}-a_{{1}{2}5} a_{{3}{4}5})\,.
 \eal
In this way $U^{1234}$ has been reduced to the desired structure.

Now we look for a similar reduction of the quantity $V^{1234}$. Here we perform a sequence of the following transformations. 
First, we have 
  \bal
 V^{1234}&=4\big(-\na_{\a} Y^{[1} \na^2 Y^{4]}\na_{\g}\na^{\a} Y^{[2} \na^{\g} Y^{3]}-\na^{\a} Y^{[1} \na^{\b} Y^{4]}\na_{\b}\na_{\a}\na_{\g} Y^{[2} \na^{\g} Y^{3]}\\
 &-\na_{\a} Y^{[1} \na_{\b} Y^{4]}\na_{\g}\na^{\a} Y^{[2} \na^{\b}\na^{\g} Y^{3]}\\
 & -\na_{\a} Y^{[1} \na^2 Y^{3]}\na_{\g}\na^{\a} Y^{[2} \na^{\g} Y^{4]}-\na^{\a} Y^{[1} \na^{\b} Y^{3]}\na_{\b}\na_{\a}\na_{\g} Y^{[2} \na^{\g} Y^{4]}\\
 &-\na_{\a} Y^{[1} \na_{\b} Y^{3]}\na_{\g}\na^{\a} Y^{[2} \na^{\b}\na^{\g} Y^{4]}\big)\, .
 \eal
 Here the combinations $\na^{\a} Y^{[1} \na^{\b} Y^{4]}$ and $\na^{\a} Y^{[1} \na^{\b} Y^{3]}$ entering in the 2nd and 5th terms are anti-symmetric in $\alpha$ and $\beta$ and, therefore, in these terms one can replace $\nabla_{\beta}\nabla_{\alpha}\nabla_{\gamma}$ with $\sfrac{1}{2}[\nabla_{\beta},\nabla_{\a}]\nabla_{\gamma}$ and then apply identity (\ref{comm}).
 In this way we get
 \bea
 V^{1234}&=&-4\big(\na_{\a} Y^{[1} \na_{\b} Y^{4]}\na_{\g}\na^{\a} Y^{[2} \na^{\b}\na^{\g} Y^{3]}+\na_{\a} Y^{[1} \na_{\b} Y^{3]}\na_{\g}\na^{\a} Y^{[2} \na^{\b}\na^{\g} Y^{4]} \nonumber \\
 &+&\na_{\a} Y^{[1} \na^2 Y^{4]}\na_{\g}\na^{\a} Y^{[2} \na^{\g} Y^{3]}-\na_{\a} Y^{[1} \na_{\b} Y^{4]}\na^{\b} Y^{[2} \na^{\a} Y^{3]}\\
 & +&\na_{\a} Y^{[1} \na^2 Y^{3]}\na_{\g}\na^{\a} Y^{[2} \na^{\g} Y^{4]}-\na_{\a} Y^{[1} \na_{\b} Y^{3]}\na^{\b} Y^{[2} \na^{\a} Y^{4]}\big)\,.  \nonumber
 \eea
 As the next step, we consider the first line in the expression above and transform it in the following way
 \bea
 \hspace{-1cm}I^{1234}&\equiv& -4\big(\na_{\a} Y^{[1} \na_{\b} Y^{4]}\na_{\g}\na^{\a} Y^{[2} \na^{\b}\na^{\g} Y^{3]}+\na_{\a} Y^{[1} \na_{\b} Y^{3]}\na_{\g}\na^{\a} Y^{[2} \na^{\b}\na^{\g} Y^{4]}\big) \nonumber \\
 &=& -(\na^{\g} Y^{1} \na^{\b} Y^{4}-\na^{\g} Y^{4} \na^{\b} Y^{1})(\na_{\g}\na_{\a} Y^{2} \na_{\b}\na^{\a} Y^{3}-\na_{\g}\na_{\a} Y^{3} \na_{\b}\na^{\a} Y^{2})\nonumber \\
 && -(\na^{\g} Y^{1} \na^{\b} Y^{3}-\na^{\g} Y^{3} \na^{\b} Y^{1})(\na_{\g}\na_{\a} Y^{2} \na_{\b}\na^{\a} Y^{4}-\na_{\g}\na_{\a} Y^{4} \na_{\b}\na^{\a} Y^{2})\nonumber \\
 &=&-2\na_{\b}\na_{\a} Y^{3}\na^{\b} Y^{4}\na_{\g}\na^{\a} Y^{2} \na^{\g} Y^{1} +2\na_{\b}\na_{\a} Y^{3} \na^{\b} Y^{1}\na_{\g}\na^{\a} Y^{2} \na^{\g} Y^{4}\nonumber \\
 &&-2 \na_{\b}\na_{\a} Y^{4}\na^{\b} Y^{3}\na_{\g}\na^{\a} Y^{2}\na^{\g} Y^{1} +2 \na_{\b}\na_{\a} Y^{4}\na^{\b} Y^{1} \na_{\g}\na^{\a} Y^{2}\na^{\g} Y^{3}\, .
 \eea
 The resulting expression undergoes further transformation 
 \bea
 &&I^{1234}=-2\na_{\a}(\na_{\b} Y^{3}\na^{\b} Y^{4})\na_{\g}\na^{\a} Y^{2} \na^{\g} Y^{1} \\
 &&~+2\big(\na_{\b}\na_{\a} Y^{[3} \na^{\b} Y^{1]}+{1\ov2}\na_{\a}(\na_{\b} Y^{3} \na^{\b} Y^{1})\big)\big(\na_{\g}\na_{\a} Y^{[2} \na^{\g} Y^{4]}+{1\ov2}\na_{\a}(\na_{\g} Y^{2} \na^{\g} Y^{4}) \big)\nonumber\\
 &&~+2\big(\na_{\b}\na_{\a} Y^{[4} \na_{\b} Y^{1]}+{1\ov2}\na_{\a}(\na_{\b} Y^{4} \na_{\b} Y^{1})\big)\big(\na_{\g}\na^{\a} Y^{[2} \na^{\g} Y^{3]}+{1\ov2}\na^{\a}(\na_{\g} Y^{2} \na^{\g} Y^{3}) \big)\, ,
 \nonumber
 \eea
 which finally results into
 \bea
 I^{1234}&=-&2\big(\na_{\b}\na_{\a} Y^{[1} \na^{\b} Y^{4]}\na_{\g}\na^{\a} Y^{[2} \na^{\g} Y^{3]} +\na_{\b}\na_{\a} Y^{[1} \na^{\b} Y^{3]}\na_{\g}\na^{\a} Y^{[2} \na^{\g} Y^{4]}\big)\nonumber \\
 &&- \na_{\b}\na_{\a} Y^{[1} \na^{\b} Y^{3]}\na^{\a}(\na_{\g} Y^{2} \na^{\g} Y^{4}) +\na_{\a}(\na_{\b} Y^{3} \na^{\b} Y^{1})\na_{\g}\na^{\a} Y^{[2} \na^{\g} Y^{4]} \nonumber \\
 &&+{1\ov2}\na_{\a}(\na_{\b} Y^{3} \na^{\b} Y^{1})\na^{\a}(\na_{\g} Y^{2} \na^{\g} Y^{4}) \\
 &&- \na_{\b}\na_{\a} Y^{[1} \na^{\b} Y^{4]}\na^{\a}(\na_{\g} Y^{2} \na^{\g} Y^{3}) +\na_{\a}(\na_{\b} Y^{4} \na^{\b} Y^{1})\na^{\g}\na^{\a} Y^{[2} \na_{\g} Y^{3]} \nonumber \\
 &&+{1\ov2}\na_{\a}(\na_{\b} Y^{4} \na^{\b} Y^{1})\na^{\a}(\na_{\g} Y^{2} \na^{\g} Y^{3}) -2\na_{\a}(\na_{\b} Y^{3}\na^{\b} Y^{4})\na^{\g}\na^{\a} Y^{2} \na_{\g} Y^{1} \,. \nonumber
  \eea
 Comparing the first line in the above formula with the original expression (\ref{V1234}) for $V^{1234}$ we observe that it coincides with $-\sfrac{1}{2}V^{1234}$. This allows us to find
 the following answer for $V^{1234}$ 
  \bea
V^{1234}&=& {2\ov3}\Big(\na^2\na_{\b} Y^{[1} \na^{\b} Y^{3]}\na_{\g} Y^{2} \na^{\g} Y^{4} -\na_{\b} Y^{3} \na^{\b} Y^{1}\na^2\na_{\g} Y^{[2} \na^{\g} Y^{4]} \nonumber \\
 &&~-{1\ov2}\na_{\b} Y^{3} \na^{\b} Y^{1}\na^2(\na_{\g} Y^{2} \na^{\g} Y^{4}) -{1\ov2}\na_{\b} Y^{4} \na^{\b} Y^{1}\na^2(\na_{\g} Y^{2} \na^{\g} Y^{3}) \nonumber \\
 &&~+\na^2\na_{\b} Y^{[1} \na^{\b} Y^{4]}\na_{\g} Y^{2} \na^{\g} Y^{3} -\na_{\b} Y^{4} \na^{\b} Y^{1}\na^2\na_{\g} Y^{[2} \na^{\g} Y^{3]} \\
 &&~+ 2\na_{\b} Y^{3}\na^{\b} Y^{4}\na_{\g}\na_{\a} Y^{2} \na^{\a} \na^{\g} Y^{1}+2\na_{\b} Y^{3}\na^{\b} Y^{4}\na^2\na_{\g} Y^{2} \na^{\g} Y^{1} \nonumber\\
 &&~-4\na_{\a} Y^{[1} \na^2 Y^{4]}\na^{\g}\na^{\a} Y^{[2} \na_{\g} Y^{3]}+4\na_{\a} Y^{[1} \na_{\b} Y^{4]}\na^{\b} Y^{[2} \na^{\a} Y^{3]} \nonumber\\
 &&~-4\na_{\a} Y^{[1} \na^2 Y^{3]}\na^{\g}\na^{\a} Y^{[2} \na_{\g} Y^{4]}+4\na_{\a} Y^{[1} \na_{\b} Y^{3]}\na^{\b} Y^{[2} \na^{\a} Y^{4]}\Big)\,. \nonumber
 \eea
 All the terms in the right hand side of the last formula are reducible, {\it i.e.} by using eqs.(\ref{Laplacian}), (\ref{rel1}), (\ref{bc}), (\ref{LspY}) they can be written via $f_5^n a_{125}a_{345}$ and permutations thereof. For instance,
 \bea\nonumber
\na^2\na_{\b} Y^{[1} \na^{\b} Y^{3]}\na_{\g} Y^{2} \na^{\g} Y^{4}&=&{1\ov2}(-f_1\na_{\b} Y^{1} \na^{\b} Y^{3}+f_3\na_{\b} Y^{3} \na^{\b} Y^{1})\na_{\g} Y^{2} \na^{\g} Y^{4}
\\
& =&{1\ov2}(f_3-f_1)b_{135}b_{245}\, .
\eea
Proceeding in a similar manner, after tedious computation we find 
 \bal
 V^{1234}&=\frac{1}{6} a_{125} a_{345} \Big(-(f_3+f_4-f_5) f_1^2+(-f_3^2+(2
   (f_2+f_4-2)+f_5) f_3\\
   &-(-2 f_2+f_4+4) (f_4-f_5))
   f_1-(f_2-f_5) (f_3^2+(f_2-2 f_4+4) f_3
   \\
   &+(f_4-f_5)
   (f_2+f_4+f_5+4))\Big)\\
   &+\frac{1}{12} a_{145} a_{235}
   \Big((f_2+f_3-f_5) f_1^2+(f_2^2-(2 (f_3+f_4-2)+f_5)
   f_2\\
   &+(f_3-2 f_4+4) (f_3-f_5)) f_1+(f_4-f_5)
   (f_2^2+(-2 f_3+f_4+4) f_2\\
   &+(f_3-f_5)
   (f_3+f_4+f_5+4))\Big)\\
   &+\frac{1}{12} a_{135} a_{245}
   \Big((f_2+f_4-f_5) f_1^2+(f_2^2-(2 (f_3+f_4-2)+f_5)
   f_2\\
   &+(-2 f_3+f_4+4) (f_4-f_5)) f_1+(f_3-f_5)
   (f_2^2+(f_3-2 f_4+4) f_2\\
   &+(f_4-f_5)
   (f_3+f_4+f_5+4))\Big) \, ,
   \eal
which according to eq.(\ref{K1234}) gives the final result for $K^{1234}$
 {\small
 \bea
&&K^{1234}=  \nonumber \\
   && \frac{1}{24} a_{125} a_{345} \Big(-4 f_5^3+4 (3 f_3+3 f_4-28) f_5^2-4 (2 f_4^2+5
   f_3 f_4-28 f_4+2 (f_3-14) f_3+48) f_5 \nonumber\\
   &&+6 (f_3-4) (f_4-4)
   (f_3+f_4)+f_2 (5 f_3^2+8 f_4 f_3-64 f_3+5 f_4^2+12 f_5^2-64 f_4 \nonumber\\
   &&-14
   (f_3+f_4-8) f_5+96)+f_1 (6 f_2^2+4 (2 (f_3+f_4-6)-5
   f_5) f_2+5 f_3^2+5 f_4^2+12 f_5^2-64 f_3 \nonumber\\
   &&+8 f_3 f_4-64 f_4-14 (f_3+f_4-8)
   f_5+96)+f_2^2 (5 f_3+5 f_4-8 (f_5+3)) \nonumber\\
   &&+f_1^2 (6 f_2+5 f_3+5
   f_4-8 (f_5+3))\Big) \nonumber\\
   &&+\frac{1}{24} a_{145} a_{235} \Big(6
   (-f_2-f_4+4) ((f_2-f_5) (f_4-f_5)-f_1 f_3) \nonumber\\
   &&-3
   (f_1+f_3-4) (f_2+f_4-4) (f_1+f_2+f_3+f_4-2 f_5)-8 f_1 f_4
   (f_2+f_3-f_5) \nonumber\\
   &&-8 f_2 f_3 (f_1+f_4-f_5)+2 (f_2-f_3)
   (f_2+f_3-f_5) (f_1+f_4-f_5) \nonumber\\
   &&+4 (f_1+f_3)
   (f_2+f_3-f_5) (f_1+f_4-f_5)+2 (f_4-f_1)
   (f_2+f_3-f_5) (f_1+f_4-f_5) \nonumber\\
   &&+8 (f_2+f_3-f_5)
   (f_1+f_4-f_5)+2 (f_2+f_3-f_5) (f_1+f_4-f_5) f_5 \nonumber\\
   &&+6
   (-f_1-f_3+4) ((f_5-f_1) (f_5-f_3)-f_2
   f_4)\Big) \nonumber\\
   &&+\frac{1}{24} a_{135} a_{245} \Big(6 (-f_1-f_4+4)
   ((f_1-f_5) (f_4-f_5)-f_2 f_3) \nonumber\\
   &&-3 (f_2+f_3-4)
   (f_1+f_4-4) (f_1+f_2+f_3+f_4-2 f_5)-8 f_2 f_4 (f_1+f_3-f_5) \nonumber\\
   &&-8
   f_1 f_3 (f_2+f_4-f_5)+2 (f_3-f_1) (f_1+f_3-f_5)
   (f_2+f_4-f_5) \nonumber\\
   &&+2 (f_2-f_4) (f_1+f_3-f_5)
   (f_2+f_4-f_5)+4 (f_1+f_4) (f_1+f_3-f_5)
   (f_2+f_4-f_5) \nonumber\\
   &&+8 (f_1+f_3-f_5) (f_2+f_4-f_5)+2
   (f_1+f_3-f_5) (f_2+f_4-f_5) f_5 \nonumber\\
   &&+6 (-f_2-f_3+4)
   ((f_5-f_2) (f_5-f_3)-f_1 f_4)\Big)\,.
 \eea
 }


\begin{thebibliography}{}

\bibitem{Mack:2009mi}
  G.~Mack,
  ``D-independent representation of Conformal Field Theories in D dimensions via transformation to auxiliary Dual Resonance Models. Scalar amplitudes,''
  arXiv:0907.2407 [hep-th].
  
\bibitem{Penedones:2010ue}
  J.~Penedones,
  ``Writing CFT correlation functions as AdS scattering amplitudes,''
  JHEP {\bf 1103} (2011) 025
  doi:10.1007/JHEP03(2011)025
  [arXiv:1011.1485 [hep-th]].
  
\bibitem{Fitzpatrick:2011ia}
  A.~L.~Fitzpatrick, J.~Kaplan, J.~Penedones, S.~Raju and B.~C.~van Rees,
  ``A Natural Language for AdS/CFT Correlators,''
  JHEP {\bf 1111} (2011) 095
  doi:10.1007/JHEP11(2011)095
  [arXiv:1107.1499 [hep-th]].
  
\bibitem{Rastelli:2016nze}
  L.~Rastelli and X.~Zhou,
  ``Mellin amplitudes for $AdS_5\times S^5$,''
  arXiv:1608.06624 [hep-th].
  

  
\bibitem{Arutyunov:1999fb}
  G.~Arutyunov and S.~Frolov,
  ``Scalar quartic couplings in type IIB supergravity on ${\rm AdS}_5\times {\rm S}^5$,''
  Nucl.\ Phys.\ B {\bf 579} (2000) 117
  doi:10.1016/S0550-3213(00)00210-8
  [hep-th/9912210].
  
\bibitem{Arutyunov:2002fh}
  G.~Arutyunov, F.~A.~Dolan, H.~Osborn and E.~Sokatchev,
  ``Correlation functions and massive Kaluza-Klein modes in the AdS/CFT correspondence,''
  Nucl.\ Phys.\ B {\bf 665} (2003) 273
  doi:10.1016/S0550-3213(03)00448-6
  [hep-th/0212116].

  

\bibitem{Arutyunov:2000py}
  G.~Arutyunov and S.~Frolov,
  ``Four point functions of lowest weight CPOs in N=4 SYM(4) in supergravity approximation,''
  Phys.\ Rev.\ D {\bf 62} (2000) 064016
  doi:10.1103/PhysRevD.62.064016
  [hep-th/0002170].
  
   
\bibitem{Arutyunov:2003ae}
  G.~Arutyunov and E.~Sokatchev,
  ``On a large N degeneracy in N=4 SYM and the AdS/CFT correspondence,''
  Nucl.\ Phys.\ B {\bf 663} (2003) 163
  doi:10.1016/S0550-3213(03)00353-5
  [hep-th/0301058].
  
\bibitem{Berdichevsky:2007xd}
  L.~Berdichevsky and P.~Naaijkens,
  ``Four-point functions of different-weight operators in the AdS/CFT correspondence,''
  JHEP {\bf 0801} (2008) 071
  doi:10.1088/1126-6708/2008/01/071
  [arXiv:0709.1365 [hep-th]].

\bibitem{Uruchurtu:2008kp}
  L.~I.~Uruchurtu,
  ``Four-point correlators with higher weight superconformal primaries in the AdS/CFT Correspondence,''
  JHEP {\bf 0903} (2009) 133
  doi:10.1088/1126-6708/2009/03/133
  [arXiv:0811.2320 [hep-th]].

\bibitem{Uruchurtu:2011wh}
  L.~I.~Uruchurtu,
  ``Next-next-to-extremal Four Point Functions of N=4 1/2 BPS Operators in the AdS/CFT Correspondence,''
  JHEP {\bf 1108} (2011) 133
  doi:10.1007/JHEP08(2011)133
  [arXiv:1106.0630 [hep-th]].


\bibitem{Arutyunov:2000ima}
  G.~Arutyunov and S.~Frolov,
  ``On the correspondence between gravity fields and CFT operators,''
  JHEP {\bf 0004} (2000) 017
  doi:10.1088/1126-6708/2000/04/017
  [hep-th/0003038].
  
   

\end{thebibliography}
\end{document}